\documentclass[12pt,smallcaptions,smallabstract]{dccpaper}
\usepackage{times,amsmath,amssymb,graphicx,booktabs,array,url,cite,microtype}
\expandafter\def\expandafter\UrlBreaks\expandafter{\UrlBreaks\do\_\do\-}
\graphicspath{{figures/}}
\title{Cartesian Byte-Bank Models and Warp Lag Screening\\for GPU Lossless Compression}
\author{Zeyu Jia $^{1,2}$\\[0.15 em]
\normalsize $^{1}$ School of Biomedical Engineering \& Technology, Tianjin Medical Univ., China\\
\normalsize $^{2}$ Medical School, Tianjin University, China (\texttt{jiazeyu@tju.edu.cn})}
\date{}

\begin{document}
\maketitle
\begin{abstract}
Headerless multi-channel 16-bit streams degrade unit-lag differencing, byte shuffling, and chunked GPU codecs when adjacent words belong to distinct sensors. On even-length 16-bit tiles ($\le 16$\,MiB), GPX6 pairs a 32-lane CUDA warp lag screen with a threshold-free $3\times 2$ Cartesian conditional byte-bank range Asymmetric Numeral Systems (rANS) coder and OpenMP Zstandard-1 framing. Across 30 subject-disjoint waveform files (88.90\,MB, $N=27$ subjects), GPX6 compresses to 26,449,754\,B (29.75\% aggregate; 42.68\% subject mean; 40.74\% six-cohort mean across 338.25\,MB) at 424.1/279.0\,MB/s host-to-host on RTX~4080 Laptop GPU and 266.9/166.4\,MB/s on A100-80GB, reaching 474.1/482.3\,MB/s (4.81/5.98\,GB/s resident) on 30-tile A100 batches. Completing GPX4 into GPX6 saves $+2.59$\,pp subject mean, localized to step-quantized BIDMC telemetry ($+6.24$\,pp vs.\ $+0.20$\,pp on smooth cohorts). Under 1-payload screening, GPX6 saves $+10.59$\,pp ($[+8.09, +13.28]$\,pp) over faster zero-order GPU \texttt{nvCOMP ANS}, beats 16-bit \texttt{MANS} and 64\,KiB \texttt{nvCOMP Zstd} ($2.17\times$--$2.49\times$ faster encode), and maps CPU, audio, and 2D boundaries.
\end{abstract}

\section{Introduction}
Serialized probability tables, host--device copies, chunk directories, and verified recovery often dominate GPU compression latency; evaluation thus requires host-to-host throughput (\emph{including copies, framing, and checks}) alongside resident kernel rates and wire bytes across matched GPU and CPU baselines. Headerless little-endian 16-bit payloads arise in Direct Memory Access (DMA) ring buffers, telemetry packets, stripped \texttt{.dat} files, and database pages decoupled from metadata. Unit-stride arrays ($s_{\text{true}}=1$) correlate across adjacent words, whereas interleaved $C$-channel streams repeat each sensor every $C$ words, inflating unit-lag differencing and shuffling when $C$ is unknown.

Relative to a four-mode split-bank baseline (GPX4: Modes~0--2, 4 under $\tau^*$ gating), GPX6 pairs two separable GPU mechanisms. First, a 32-lane CUDA warp screens lags $s \in \{1,\dots,32\}$ in $22$--$91\,\mu\text{s}$/tile and feeds $s_{\min}$ to a threshold-free six-mode Minimum Description Length selector (\emph{Full-6 MDL}, zero empirical thresholds). Second, at fixed two-bank capacity ($2\times 256\times 256$ uint16, 256\,KiB raw), GPX6 completes the $3\times 2$ Cartesian product of three word transforms and two byte conditionings via fused GPU histogramming, OpenMP parallel Zstandard-1 (Zstd-1) framing, split-bank GPU rANS kernels, and a CUDA-free C++ \texttt{.dll}/\texttt{.so} receiver. Cross-plugging our screen across external back-ends separates both pillars: screening discovers favorable lags (including harmonic $s^*=8$ on $C=2$ Fantasia) without headers. Under matched 1-payload selection, Cartesian rANS saves $7.10$--$10.72$\,pp over faster zero-order GPU \texttt{nvCOMP ANS}, beats 16-bit GPU/CPU \texttt{MANS}~\cite{mans2025primary} and 64\,KiB \texttt{nvCOMP Zstd}, and cuts step-quantized intensive-care records by $6.24$\,pp over GPX4. GPX6 requires even byte alignment ($n=2W$), screens $s \le 32$ autonomously (external strides $\le 1{,}024$), and targets 16-bit words.

\section{Related Work}
Asymmetric Numeral Systems (ANS)~\cite{duda2013ans} and interleaved ANS~\cite{giesen2014interleaved} pair near-entropy rates with SIMD/SIMT lookups. GPU entropy coders target zero-order speed, range mapping, or chunked Lempel--Ziv (LZ): \texttt{multians}~\cite{weissenberger2019multians} merges zero-order states, Recoil~\cite{recoil2023} scales adaptive rANS decoding, and DietGPU~\cite{johnson2022dietgpu} and \texttt{nvCOMP}~\cite{nvidia2023nvcomp} provide chunked zero-order ANS, LZ/Zstd, Bitcomp, and Cascaded kernels. MANS~\cite{mans2025primary} pairs 1D/3D Adaptive Data Mapping (ADM) with GPU P-mode ANS and CPU R-mode FSE on 16-bit integers, \texttt{ndzip-gpu}~\cite{knorr2021ndzipgpu} compacts Lorenzo bit planes, and LC~\cite{azami2025lc} searches primitive pipelines offline. Within our split-bank family, GPX6 isolates residual high-first conditioning (Modes~3, 5), Full-6 MDL, fused histogramming, and parallel framing against GPX4.

On CPUs, Zstd~\cite{collet2021rfc8878} pairs LZ77 with FSE, and Blosc2~\cite{alted2023blosc2} shuffles byte/bit planes given stride. PAQ8's \texttt{RecordModel}~\cite{mahoney2005paq} detects byte record lengths for serial arithmetic coding, whereas GPX6 screens 32 word lags in one CUDA warp via \texttt{\_\_clz}. Rice~\cite{rice1979}, JPEG~\cite{wallace1991jpeg}, and Pcodec~\cite{loncaric2025pcodec} emit unconditioned offset bits inside bins; GPX6 splits 16-bit words at byte boundaries and conditions $L'_w$ on $H'_w$ ($256\times 256$ lookup) to capture quantized analog-to-digital converter (ADC) steps and sign folds in parallel. OpenZL~\cite{openzlgraphs2026} compiles typed graphs given stride, FLAC~\cite{vanbeurden2024flac} caps linear predictive coding (LPC) at eight channels, MPEG-4 ALS~\cite{kamamoto2008mpeg4als} extends joint LPC given header $C$, and LICO~\cite{azami2024lico}, JPEG-LS~\cite{weinberger2000locoi}, HTJ2K~\cite{taubman2022htj2k}, JPEG~XL~\cite{alakuijala2019jpegxl}, and probabilistic circuits~\cite{severo2025pc} require 2D $(H,W)$. Following two-pass MDL~\cite{krishnan2014universal,damme2019selection}, we test GPX6 against $s=1$, warp-screened (1-payload $\tau^*$ and 2-payload MDL), \emph{Oracle-$C$}, \texttt{MANS}, and 2D codecs.

\section{Cartesian Byte-Bank Models, Full-6 MDL, and GPU Kernels}
For $W = n/2$ little-endian 16-bit words $x_w = L_w + 256 H_w$ ($2 \le n \le 16\text{\,MiB}$) and fixed two-bank capacity ($2\times 256\times 256$ uint16, 256\,KiB raw), GPX6 crosses three word transforms with two byte conditionings (Table~\ref{tab:dependencies}, Figure~\ref{fig:architecture}; magics in Supplement~S1). Modular differences $d_w = (u_w - u_{w-1}) \bmod 2^{16}$ ($u_{-1}=0$) map bijectively to zigzag residuals $z_w = \mathrm{zz}(d_w) = ((d_w \ll 1) \oplus (-(d_w \gg 15))) \bmod 2^{16} = L'_w + 256 H'_w$; Modes~3 and~5 condition $L'_w$ on $H'_w$ rather than $L'_{w-1}$ across quantized ADC grids.

\begin{figure}[!htbp]
\centering
\includegraphics[width=0.96\linewidth]{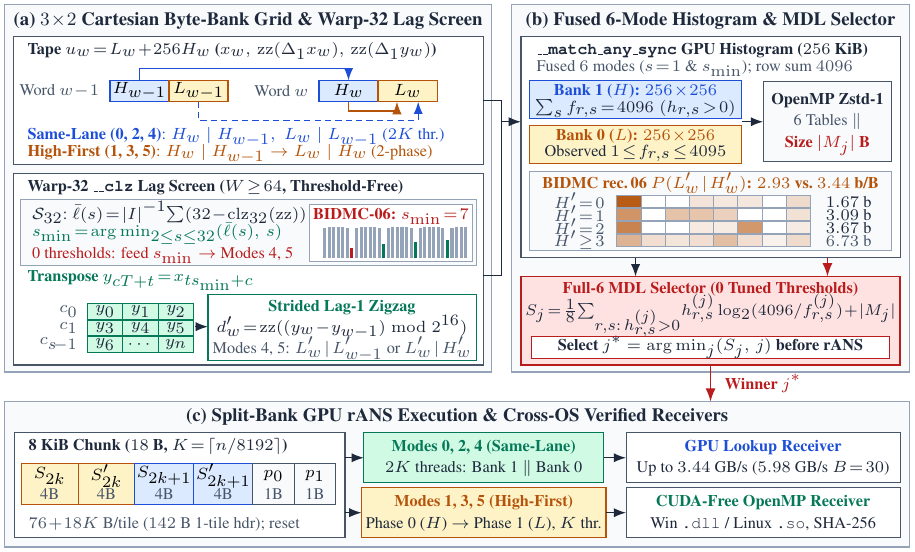}
\caption{GPX6 architecture: (a) $3\times 2$ byte banks and warp lag screen, (b) fused histogram and Full-6 MDL, (c) GPU/CPU rANS.}
\label{fig:architecture}
\end{figure}

For $W \ge 64$, one 32-lane CUDA warp screens $s \in \{1,\dots,32\}$ on $I = \{32 + m\Delta : 0 \le m < |I|\}$, $\Delta = \max(1, \lfloor(W-32)/65536\rfloor)$, $|I| = \min(65536, W-32)$, summing $32 - \mathrm{clz}_{32}(\mathrm{zz}((x_w - x_{w-s})\bmod 2^{16}))$:
\begin{equation}
\bar{\ell}(s) = \frac{1}{|I|}\sum_{w \in I} \bigl\lceil \log_2\!\bigl(1 + \mathrm{zz}((x_w - x_{w-s})\bmod 2^{16})\bigr)\bigr\rceil, \quad
y_{cT + t} = x_{t s_{\text{cand}} + c},
\label{eq:screen}
\end{equation}
for $0 \le t < T = \lfloor W / s_{\text{cand}}\rfloor$ and $0 \le c < s_{\text{cand}}$, with tail identity $y_w = x_w$ for $T s_{\text{cand}} \le w < W$ (or per 4,096-word block under \texttt{chunk\_reset}). With $s_{\min} = \arg\min_{2 \le s \le 32:\, n \ge 2s}(\bar{\ell}(s), s)$, default \emph{Full-6 MDL} sets $s_{\text{cand}} = s_{\min}$ if $W \ge 64$ and $\bar{\ell}(1) > 0$ ($1$ otherwise; direct strided calls use $\min(12, W)$), scoring all $k_{\text{act}}=6$ modes with zero empirical thresholds; calibration $\tau^* = \sqrt{r_{\text{multi}}^{\max} r_{\text{single}}^{\min}} = 0.970676$ skips Modes~4--5 when $\bar{\ell}(s_{\min})/\bar{\ell}(1) \ge \tau^*$.

\begin{table}[!htbp]
\centering
\scriptsize
\renewcommand{\arraystretch}{0.76}
\setlength{\tabcolsep}{4.4pt}
\begin{tabular}{lllll}\toprule
Mode Index and Semantic Name & Word Transform $u_w$ & Bank 1 ($H$) & Bank 0 ($L$) & Decode Schedule \\\midrule
Mode 0: Identity same-lane & Identity $x_w$ & $H_w \mid H_{w-1}$ & $L_w \mid L_{w-1}$ & $2K$ concurrent \\
Mode 1: Identity high-first & Identity $(H_w, L_w)$ & $H_w \mid H_{w-1}$ & $L_w \mid H_w$ & 2-phase ($K\to K$) \\
Mode 2: Residual same-lane & $\mathrm{zz}(\Delta_1 x_w)$ ($s^*=1$) & $H'_w \mid H'_{w-1}$ & $L'_w \mid L'_{w-1}$ & $2K$ concurrent \\
Mode 3: Residual high-first & $\mathrm{zz}(\Delta_1 x_w)$ ($s^*=1$) & $H'_w \mid H'_{w-1}$ & $L'_w \mid H'_w$ & 2-phase ($K\to K$) \\
Mode 4: Strided res.\ same-lane & Transpose $s_{\text{cand}}\!\to\!\mathrm{zz}(\Delta_1 y_w)$ & $H'_w \mid H'_{w-1}$ & $L'_w \mid L'_{w-1}$ & $2K$ concurrent \\
Mode 5: Strided res.\ high-first & Transpose $s_{\text{cand}}\!\to\!\mathrm{zz}(\Delta_1 y_w)$ & $H'_w \mid H'_{w-1}$ & $L'_w \mid H'_w$ & 2-phase ($K\to K$) \\\bottomrule
\end{tabular}
\caption{$3\times 2$ Cartesian 512-row byte-bank family ($2\times 256\times 256$ uint16; GPX4: 0--2, 4).}
\label{tab:dependencies}
\vspace{2pt}
\setlength{\tabcolsep}{2.6pt}
\begin{tabular}{ll}\toprule
Stage / Kernel & Parallel Mapping, Memory Traffic, and Execution Invariant \\\midrule
1. \texttt{screen\_strides} & Grid $\min(256,\lceil|I|/8\rceil)\!\times\!256$; lane $s\!-\!1$ sums $32\!-\!\mathrm{clz}_{32}(\mathrm{zz}(\Delta_s x_w))$; transpose $s_{\min}$. \\
2. \texttt{hist\_cartesian6} & Fused 1-thread/byte pass on $i \in [2,n)$, $(i \bmod 8192)\!\ge\!2$ over $x,y$ via \texttt{\_\_match\_any\_sync}. \\
3. \texttt{normalize}\,+\,OMP & GPU normalizes $512 k_{\text{act}}$ rows to 4096 ($f^{(j)}_{r,s}\!\in\![1,4095]$, $h^{(j)}_{r,s}\!>\!0$); OMP Zstd-1 picks $j^*$. \\
4. \texttt{lanes\_enc / inv} & Encode: $2K$ threads ($K\!=\!\lceil n/8192\rceil$) code 4\,KiB banks in reverse; LUT decode + SHA-256. \\\bottomrule
\end{tabular}
\caption{GPX6 GPU kernel pipeline and OpenMP parallel Zstd-1 table framing schedule.}
\label{tab:pseudocode}
\end{table}

\subsection{GPU Kernel Schedule, Parallel Table Framing, and Word-Aligned Bounds}
Observed symbols ($h^{(j)}_{r,s} > 0$) in row $r \in \{0,\dots,511\}$ receive $f^{(j)}_{r,s} \in [1, 4095]$ summing to $4096$ via largest-remainder rounding ($f^{(j)}_{r,s}=16$ on empty rows; singleton rows reserve $1$ for the first absent symbol, and unobserved symbols in multi-symbol rows get $f^{(j)}_{r,s}=0$; Table~\ref{tab:pseudocode}). OpenMP C++ (\texttt{zstd\_compress\_tables\_mt}) frames all $k_{\text{act}}$ 256\,KiB tables at Zstd-1 into $M_j$ ($1.04$--$5.09$\,ms on i9-13900HX, $0.16$--$1.89$\,ms on Xeon on $\ge 7$\,MB tiles; $1.46\times$--$4.81\times$ faster than serial Python with identical bytes). MDL picks $j^* = \arg\min_j (S_j, j)$ via $S_j = \frac{1}{8}\sum_{r=0}^{511}\sum_{s:\, h^{(j)}_{r,s}>0} h^{(j)}_{r,s}\log_2(4096 / f^{(j)}_{r,s}) + |M_j|$ before coding one rANS payload ($s^* = s_{\text{cand}}$ if $j^* \in \{4,5\}$ else $1$). It incurs $0$\,B wire regret across all 72 multi-MiB tiles (432 frames; 81 tiles / 486 frames with Suite~4) and $0$\,B gap vs.\ an exhaustive $1..32$ six-mode oracle on Suite~1 (Supplement~S5).

\textbf{Lemma 1 (Word-aligned 16-bit rANS frame bound).} \emph{For $n=2W \in [2, 16\text{\,MiB}]$, $K = \lceil n/8192\rceil$, $\sum_{r,s:\,h_{r,s}>0} h_{r,s} = \sum_{k=0}^{K-1}\max(0, \ell_k - 2)$, and observed $(r,s)$ with $h_{r,s}>0$, $f_{r,s} \in [1, 4095]$ ($\sum_{s=0}^{255} f_{r,s} = 4096$, $c_{r,s} = \sum_{u<s} f_{r,u} \le 4095$), $x_0 = 65536$, and $B_{r,s} = 2^{20} f_{r,s} \ge 2^{20} > 65535 = \lfloor(2^{32}-1)/2^{16}\rfloor$, 16-bit shifts emit $w_t \in \{0,1\}$ uint16 words per coded byte ($x_t \in [65536, 2^{32}-1]$), bounding single-tile wire bytes:}
\begin{equation}
|W_{\text{out}}| \le 142\!+\!|M_{j^*}|\!+\!18K\!+\!2\!\left\lfloor \tfrac{1}{16}\textstyle\sum\limits_{h_{r,s}>0} h_{r,s}\bigl(12\!-\!\log_2 f_{r,s}\!+\!\log_2(1\!+\!c_{r,s}/65536)\bigr)\right\rfloor.
\label{eq:lemma}
\end{equation}
Rational interval evaluation of Eq.~\eqref{eq:lemma} encloses all 432 real mode frames across 72 tiles (max slack 83,162\,B; Supplement~S1). Suite~1 wire bytes (29.75\%) split into 28.78\% plug-in model conditional entropy, 0.06\% 12-bit KL, 0.22\% chunk records, and 0.71\% Zstd-1 tables minus 0.02\% state slack. Total framing/coding overhead is $0.98$\,pp above the selected model's empirical conditional entropy ($0.58$--$0.77$\,pp on CT/EDF vs.\ $3.94$--$5.37$\,pp tables on Suite~4). Without raw passthrough, \texttt{.gz} archives expand to 100.69\%, 107.74\%, 108.93\%, and 118.95\%.

\section{Experimental Protocol}
We evaluate 70 public 16-bit files (445,996,612\,B / 446.00\,MB) across four subject- and SHA-256-disjoint splits: Development (10 files, 54.65\,MB), Calibration (15 files, 101.30\,MB, used only for $\tau^*$), multi-MiB Suites~1--3 (36 files, 287.43\,MB), and sub-MiB Suite~4 (9 files, 2.61\,MB; Supplement~S2). Given longitudinal diagnostic exposure across study iterations (Suites~2--4 in unit-stride profiling; Suite~1 in 4-vs-6-mode and $\tau^*$ checks), we report Suites~1--4 transparently as \emph{subject-disjoint reused evaluation suites}. Suite~1~\cite{pollard2026physionet} holds 30 stripped \texttt{.dat}/\texttt{.xyz} files (88.90\,MB, $N=27$ subjects; three PTB subjects pair 12-lead and $XYZ$ files averaged per subject). It spans BIDMC~\cite{pimentel2017bidmc} ($C\in\{5..7\}$, $n=9$), CEBSDB~\cite{garciagonzalez2013cebsdb} ($C=4$, $n=3$), Fantasia~\cite{iyengar1996fantasia} ($C=2$, $n=3$), GrabMyo~\cite{pradhan2022grabmyo} ($C=32$, $n=4$), and PTB~\cite{bousseljot1995ptb} ($C\in\{3,12\}$, $n=8$). Suite~2 holds three CT volumes~\cite{stanfordvolume,openscivis} (47.94\,MB), Suite~3 holds three Sleep-EDF files~\cite{kemp2000sleep} (150.60\,MB; Combined-10 adds Dev, 198.79\,MB), and Suite~4 holds three NIfTI (1.29\,MB) and six LibriSpeech clips~\cite{panayotov2015librispeech} (1.32\,MB).

After a $3.0$\,s GPU warm-up on uncontended Windows~11 (RTX~4080 Laptop, 12\,GB, 58 SMs; 20-worker i9-13900HX) and Linux (A100-SXM4-80GB, 108 SMs; 12-worker Xeon), methods and files run in randomized order over 7 repetitions (2 passes for deterministic CPU \texttt{MANS}). Cohort H2H rate is $\text{MB/s} = (\sum_{f\in\mathcal{C}} R_f / 10^6)/\operatorname{median}_{r=1..7}(\sum_{f\in\mathcal{C}} t_{f,r})$ ($10^6\,\text{B/s}$, including copies, framing, and SHA-256) alongside CUDA-event resident \texttt{GB/s}. Across 83 Suite~1 arms (Supplement~S3), we compare $s=1$ codecs, warp-screened back-ends (1-payload $\tau^*$ and 2-payload MDL-2), 8\,KiB and 64\,KiB \texttt{nvCOMP}, 16-bit \texttt{MANS} (\texttt{-u2} CPU P/R and CUDA P-mode)~\cite{mans2025primary}, \emph{Oracle-$C$} pipelines, and 2D CT codecs~\cite{weinberger2000locoi,taubman2022htj2k,alakuijala2019jpegxl}. Per-subject saving $\Delta r_i = 100(W_{i,\text{base}} - W_{i,\text{GPX6}})/R_i$ ($\Delta r_i > 0$ favors GPX6, $N=27$) uses exact two-sided Wilcoxon tests~\cite{wilcoxon1945individual} excluding exact ties (\texttt{wilcox}, $N_{\text{eff}} = N - N_0$) with Holm adjustment ($m=81$)~\cite{holm1979simple}, sign-flip permutation $\hat{p}_{\text{perm}} = (k+1)/(B+1) \ge 1/10001$ ($B=10{,}000$)~\cite{phipson2010permutation}, and paired bootstrap 95\% CIs~\cite{efron1979bootstrap} ($p_{\text{Holm}} \ge 0.05$: non-rejection).

\section{Results and Scope Calibration}
\subsection{Two-Pillar Attribution and Matched GPU/CPU Comparisons}
\begin{figure}[!htbp]
\centering
\includegraphics[width=0.94\linewidth]{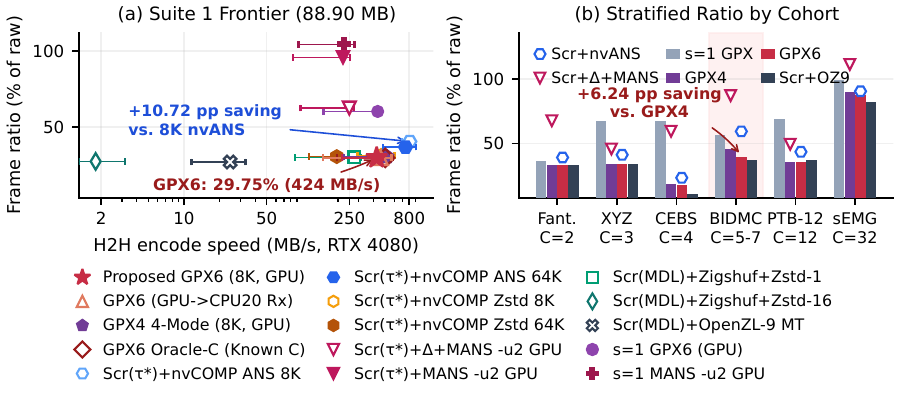}
\caption{Suite~1 ($N=27$): (a) RTX~4080 H2H encode frontier, and (b) stratified ratios across six WFDB cohorts.}
\label{fig:wfdb}
\end{figure}

On Suite~1 (Table~\ref{tab:wfdb_main}, Figure~\ref{fig:wfdb}), GPX6 compresses 88.90\,MB to 26,449,754\,B (29.75\% aggregate; 42.68\% subject mean; 37.68\% median, IQR $[33.33\%, 41.52\%]$; 40.74\% six-cohort mean vs.\ 46.55\% OpenZL-9 and 47.53\% 1-payload \texttt{nvCOMP ANS}). Steady-state H2H speed is 424.1/279.0\,MB/s on RTX~4080 (IQR $[332.5, 446.2]/[226.7, 288.0]$) and 266.9/166.4\,MB/s on A100 (IQR $[264.0, 273.5]/[162.0, 168.9]$), with identical bytes on the CUDA-free C++ receiver (215.3 and 86.0\,MB/s decode). Stride-blind ($s=1$) codecs stall at 46.79\%--104.34\% ($N_{\text{eff}}=27$, $p_{\text{Holm}}=1.21\times 10^{-6}$, $\hat{p}_{\text{perm}}=1/10001$). In 16-bit \texttt{MANS}~\cite{mans2025primary}, 512-word spans above ADM's 3,500-count ceiling emit raw 16-bit blocks. Even under 1-payload screening plus $\Delta_{s^*}$, \texttt{MANS -u2} GPU reaches 62.46\% ($\Delta r = +31.66$\,pp, $249.2/168.4$\,MB/s on 4080; 95.80\% without $\Delta_{s^*}$, 119.32\% CPU-R at $s=1$). Under 1-payload $\tau^*$ selection, zero-order GPU \texttt{nvCOMP ANS} encodes faster on 4080 (808.2\,MB/s at 8\,KiB; 743.5\,MB/s at 64\,KiB) and comparably on A100 (252.9--258.9\,MB/s). GPX6 saves $10.72$\,pp aggregate at 8\,KiB ($\Delta r = +12.76$\,pp) and $7.10$\,pp at 64\,KiB ($\Delta r = +10.59$\,pp, 95\% CI $[+8.09, +13.28]$\,pp, $p_{\text{Holm}}=1.21\times 10^{-6}$). Against 1-payload \texttt{nvCOMP Zstd}, GPX6 beats 64\,KiB chunks in ratio ($\Delta r = +2.22$\,pp, $[+1.05, +3.42]$\,pp, $p_{\text{Holm}}=0.029$) and encode speed ($2.17\times$ on 4080, $2.49\times$ on A100; $3.53\times$--$3.76\times$ vs.\ MDL-2). At 8\,KiB, \texttt{nvCOMP Zstd} reaches 463.9\,MB/s on 4080 (217.3\,MB/s on A100) while widening GPX6's saving to $+2.85$\,pp ($p_{\text{Holm}}=0.005$). Against \emph{Oracle-$C$} GPX6 (29.76\%, $N_{\text{eff}}=1$, 26 ties, $p_{\text{Holm}}=1.00$), GPX6 saves 3,373\,B on \texttt{fantasia:f1y01} via harmonic $s^*=8$ ($W \bmod 8 = 6$).

\begin{table}[!htbp]
\centering
\scriptsize
\renewcommand{\arraystretch}{0.75}
\setlength{\tabcolsep}{2.2pt}
\begin{tabular}{lrrrrrrrr}\toprule
Pipeline / Baseline Tier & Agg\% & SMn\% & Med\% & P90\% & $\Delta r$\,(pp) & $p_{\text{Holm}}$ & 4080 E/D & A100 E/D \\\midrule
\textbf{GPX6 Full-6 MDL (GPU, 8K)} & \textbf{29.75} & \textbf{42.68} & \textbf{37.68} & \textbf{84.08} & --- & --- & 424.1/279.0 & 266.9/166.4 \\
GPX6 Full-6 MDL (CPU Rx) & 29.75 & 42.68 & 37.68 & 84.08 & $+0.00$ & --- & 458.7/215.3 & 257.7/86.0 \\
GPX4 Warp-32 (4-mode, 8K) & 30.38 & 45.27 & 38.00 & 86.48 & $+2.59$ & $4.58\!\times\!10^{-4}$ & 519.9/356.7 & 262.8/188.0 \\\midrule
GPX6 Unit-Lag ($s=1$) & 60.23 & 65.15 & 61.99 & 98.38 & $+22.47$ & $1.21\!\times\!10^{-6}$ & 432.9/223.0 & 251.2/137.3 \\
MANS \texttt{-u2} P-mode GPU ($s=1$) & 104.34 & 105.67 & 116.29 & 119.24 & $+62.99$ & $1.21\!\times\!10^{-6}$ & 223.2/228.5 & 158.8/96.7 \\
Blosc2 Shuf+Zstd-9 ($s=1$) & 51.88 & 65.93 & 55.98 & 92.45 & $+23.25$ & $1.21\!\times\!10^{-6}$ & 9.5/332.5 & 13.6/275.1 \\
OpenZL-9 num16 MT ($s=1$) & 64.49 & 66.18 & 63.79 & 91.78 & $+23.50$ & $1.21\!\times\!10^{-6}$ & 49.9/393.5 & 55.4/247.3 \\
nvCOMP Zstd 64\,KiB ($s=1$) & 67.02 & 75.46 & 74.65 & 96.48 & $+32.78$ & $1.21\!\times\!10^{-6}$ & 238.4/371.2 & 124.1/173.3 \\
nvCOMP ANS 64\,KiB ($s=1$) & 88.03 & 91.96 & 98.33 & 102.18 & $+49.28$ & $1.21\!\times\!10^{-6}$ & 581.6/638.2 & 226.1/241.7 \\\midrule
Scr($\tau^*$)+$\Delta_{s^*}$+MANS \texttt{-u2} GPU & 62.46 & 74.34 & 69.88 & 107.70 & $+31.66$ & $1.21\!\times\!10^{-6}$ & 249.2/168.4 & 186.5/75.3 \\
Scr($\tau^*$)+GPU $\Delta_{s^*}$+nvZstd 8K & 31.31 & 45.53 & 39.74 & 81.87 & $+2.85$ & $0.005$ & 463.9/568.2 & 217.3/256.3 \\
Scr($\tau^*$)+GPU $\Delta_{s^*}$+nvZstd 64K & 30.53 & 44.90 & 39.08 & 82.16 & $+2.22$ & $0.029$ & 195.3/398.4 & 107.2/193.0 \\
Scr(MDL-2)+GPU $\Delta_{s^*}$+nvZstd 64K & 30.49 & 44.72 & 39.08 & 82.16 & $+2.04$ & $0.11$ & 120.3/396.7 & 71.0/192.6 \\
Scr($\tau^*$)+GPU $\Delta_{s^*}$+nvANS 8K & 40.47 & 55.44 & 48.71 & 90.17 & $+12.76$ & $1.21\!\times\!10^{-6}$ & 808.2/787.5 & 252.9/279.2 \\
Scr($\tau^*$)+GPU $\Delta_{s^*}$+nvANS 64K & 36.85 & 53.27 & 49.10 & 86.89 & $+10.59$ & $1.21\!\times\!10^{-6}$ & 743.5/698.6 & 258.9/279.0 \\
Scr(MDL-2)+GPU $\Delta_{s^*}$+nvANS 64K & 36.81 & 53.08 & 49.10 & 86.89 & $+10.40$ & $1.21\!\times\!10^{-6}$ & 506.2/779.6 & 209.4/278.4 \\
Scr(MDL-2)+ZigShuf+Zstd-16 & 27.38 & 41.63 & 36.67 & 81.55 & $-1.05$ & $1.00$ & 1.8/348.7 & 3.1/182.3 \\
Scr(MDL-2)+OpenZL-9 MT & 27.00 & 40.69 & 37.84 & 78.50 & $-1.99$ & $1.00$ & 24.6/285.5 & 25.2/207.8 \\\midrule
Oracle-$C$ GPX6 (GPU, 8K) & 29.76 & 42.68 & 37.68 & 84.08 & $+0.00$ & $1.00$ & 501.2/295.4 & 271.0/165.8 \\
Oracle-$C$ FLAC-16 LPC MT & 38.45 & 50.57 & 42.21 & 71.18 & $+7.89$ & $1.00$ & 94.4/196.0 & 107.6/151.0 \\
Oracle-$C$ GPU $\Delta_C$+nvANS 64K & 36.68 & 53.02 & 49.10 & 86.89 & $+10.34$ & $1.21\!\times\!10^{-6}$ & 826.9/757.6 & 262.8/278.0 \\\bottomrule
\end{tabular}
\caption{Suite~1 WFDB (88.90\,MB, $N=27$; 83 arms in Supp.~S3; $+\Delta r>0$ favors GPX6; $N_{\text{eff}}=27$ [GPX4 $17$, Oracle $1$]; H2H \texttt{MB/s}).}
\label{tab:wfdb_main}
\end{table}

\subsection{Orthogonal $2\times 2$ Factorials and Regime-Specific BIDMC Gain}
In Table~\ref{tab:factorials}, $\{s=1, \text{Warp-32}\} \times \{\text{GPX4}, \text{GPX6}\}$ yields 60.23\% in both arms at $s=1$ because channel mixing masks intra-word structure, and $+0.63$\,pp aggregate saving once lag is aligned ($+2.59$\,pp subject mean, $[+1.49, +3.87]$\,pp, $N_{\text{eff}}=17$, 10 ties, $\hat{p}_{\text{perm}} = 1/10001$). Table~\ref{tab:wfdb_strata} localizes this gain: on BIDMC ($n=9$, $N_{\text{eff}}=9$, $I(L'_w; H'_w) - I(L'_w; L'_{w-1}) = +0.55$\,bits/word), Mode~5 wins 9/9 files and saves $+6.24$\,pp aggregate ($+6.39$\,pp subject mean, $[+4.84, +8.29]$\,pp, $\hat{p}_{\text{perm}}=0.0044$), vs.\ $+0.20$\,pp across 18 smooth subjects ($N_{\text{eff}}=8$, $+0.69$\,pp subject mean, $[+0.25, +1.19]$\,pp). Crossing $\{\tau^*\text{-Gated}, \text{Full-6 MDL}\} \times \{\text{Serial Py}, \text{OMP C++}\}$ across 70 files verifies byte-identical OpenMP framing and saves 58,565\,B under Full-6 MDL (recovering $s^*=32$ on \texttt{grabmyo} Participant~3, $\bar{\ell}(32)/\bar{\ell}(1) = 0.9770 > \tau^*$). Capacity $512$ beats $256$ rows by $0.28$--$1.65$\,pp on multi-MiB cohorts. Under \texttt{chunk\_reset} ($\le 0.01\%$ wire cost on $s=1$, $+1.90\%$ on WFDB), cold 8\,KiB GPU recovery on 16.78\,MB \texttt{carp} takes $5.01$\,ms on 4080 ($4.69\times$ vs.\ $23.52$\,ms full tile) and $6.64$\,ms on A100 ($9.46\times$ vs.\ $62.79$\,ms). Warm LUT-cached decode drops to $2.05$\,ms ($11.46\times$) and $3.34$\,ms ($18.77\times$; Supplement~S5).

\begin{table}[!htbp]
\centering
\scriptsize
\renewcommand{\arraystretch}{0.76}
\setlength{\tabcolsep}{2.9pt}
\begin{tabular}{lrrrrrrr}\toprule
$2\times 2$ Factorial Cell & Wire\,(B) & Agg\% & SMn\% & Med\% & BIDMC\% & 4080 E/D & A100 E/D \\\midrule
\multicolumn{8}{l}{\emph{A. Cartesian Model $\times$ Lag Screen on Suite~1 (30 files, 88.90\,MB, $N=27$; BIDMC $n=9$, Smooth $n=18$)}} \\
$s=1 \times \text{GPX4}$ (4-mode) & 53,547,807 & 60.23 & 65.15 & 61.99 & 56.49 & 456.7/229.3 & 241.9/133.1 \\
$s=1 \times \text{GPX6}$ (6-mode) & 53,547,807 & 60.23 & 65.15 & 61.99 & 56.49 & 432.9/223.0 & 251.2/137.3 \\
$\text{Warp-32} \times \text{GPX4}$ (4-mode) & 27,010,139 & 30.38 & 45.27 & 38.00 & 45.72 & 519.9/356.7 & 262.8/188.0 \\
$\text{Warp-32} \times \text{GPX6}$ (proposed) & 26,449,754 & 29.75 & 42.68 & 37.68 & 39.48 & 424.1/279.0 & 266.9/166.4 \\\midrule
\multicolumn{8}{l}{\emph{B. Stride Gate $\times$ Table Framing across All 70 Files (446.00\,MB; Suite~1 steady-state E/D \texttt{MB/s} shown)}} \\
$\tau^*\text{-Gated} \times \text{Serial Python}$ & 180,858,501 & 40.55 & 43.01 & 37.68 & 39.48 & 380.9/289.9 & 226.7/158.6 \\
$\tau^*\text{-Gated} \times \text{OpenMP C++}$ & 180,858,501 & 40.55 & 43.01 & 37.68 & 39.48 & 468.9/284.1 & 265.7/159.2 \\
$\text{Full-6 MDL} \times \text{Serial Python}$ & 180,799,936 & 40.54 & 42.68 & 37.68 & 39.48 & 387.4/293.6 & 228.2/166.8 \\
$\text{Full-6 MDL} \times \text{OpenMP C++}$ & 180,799,936 & 40.54 & 42.68 & 37.68 & 39.48 & 424.1/279.0 & 266.9/166.4 \\\bottomrule
\end{tabular}
\caption{Orthogonal $2\times 2$ factorials: (A) screen $\times$ model and (B) Full-6 MDL $\times$ OpenMP Zstd-1.}
\label{tab:factorials}
\vspace{2pt}
\setlength{\tabcolsep}{2.9pt}
\begin{tabular}{lrrrrrr}\toprule
Suite~1 Cohort ($C$, subjects) & GPX6 & GPX4 & GPX6 $s\!=\!1$ & Scr($\tau^*$)+nvANS & Scr+OZ9 & Orc-FLAC \\\midrule
BIDMC ICU ($C\!\in\!\{5..7\}$, $n=9$) & 39.48 (39.66) & 45.72 (46.05) & 56.49 (56.22) & 59.48 (59.54) & 37.03 (37.19) & 67.13 (67.29) \\
CEBSDB ($C=4$, $n=3$) & 18.12 (18.12) & 18.32 (18.32) & 67.23 (67.23) & 23.47 (23.47) & 11.14 (11.14) & 39.21 (39.21) \\
Fantasia ($C=2$, $n=3$) & 33.16 (33.16) & 33.19 (33.20) & 36.42 (36.42) & 39.20 (39.21) & 33.44 (33.44) & 29.58 (29.58) \\
GrabMyo EMG ($C=32$, $n=4$) & 87.57 (87.57) & 89.77 (89.77) & 99.16 (99.16) & 90.61 (90.61) & 82.19 (82.19) & 68.84 (68.84) \\
PTB 12-Lead ($C=12$, $n=8$) & 35.84 (36.54) & 35.95 (36.89) & 68.78 (68.46) & 43.61 (44.17) & 37.33 (37.90) & 34.40 (34.98) \\
PTB $XYZ$ ($C=3$, $n=3$) & 34.05 (34.05) & 34.18 (34.18) & 67.69 (67.69) & 41.31 (41.31) & 34.25 (34.25) & 31.31 (31.31) \\\bottomrule
\end{tabular}
\caption{Stratified Suite~1 compression ratios reported as aggregate\% (subject mean\%).}
\label{tab:wfdb_strata}
\end{table}

\subsection{Stage Latencies, Batch Scaling, and Cross-Domain Boundaries}
\begin{figure}[!htbp]
\centering
\includegraphics[width=0.93\linewidth]{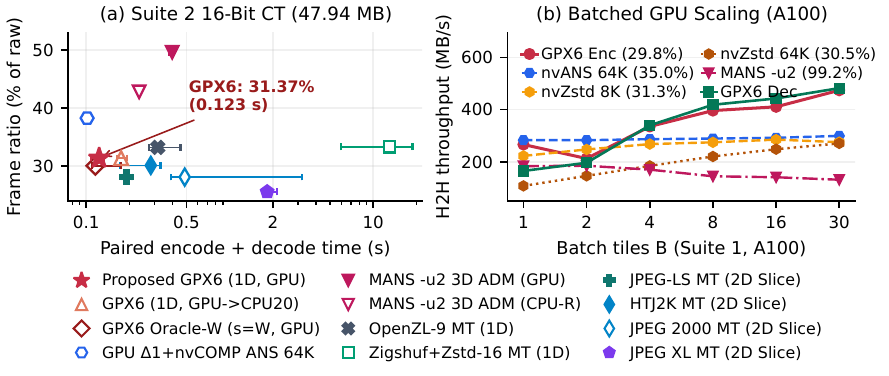}
\caption{Suite~2 CT and batch scaling: (a) 16-bit CT vs.\ 1D, 3D \texttt{MANS}, and 2D codecs, and (b) A100 batch scaling ($B=1..30$).}
\label{fig:nvcomp}
\vspace{2pt}
\includegraphics[width=0.94\linewidth]{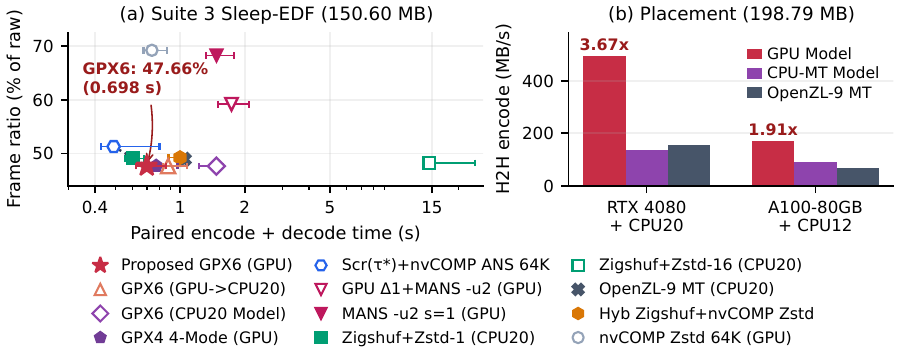}
\caption{Unit-stride frontiers: (a) Suite~3 Sleep-EDF ($s^*=1$ on 11/11 tiles) and (b) Combined-10 GPU vs.\ CPU-MT placement.}
\label{fig:edf_ct}
\end{figure}

On $\ge 12$\,MB tiles, resident GPU encode is $2.91$--$6.32$\,ms on RTX~4080 ($2.66$--$4.13$\,GB/s; frontend $0.175$--$0.205$\,ms, backend $2.73$--$6.13$\,ms; decode $2.07$--$3.36$\,GB/s). On A100, resident encode is $2.60$--$6.70$\,ms ($2.51$--$4.61$\,GB/s; frontend $0.068$--$0.094$\,ms, backend $2.54$--$6.61$\,ms; decode $1.58$--$3.44$\,GB/s; Supp.~S5d). Single sub-MiB tiles ($2K = 160$--$206$ threads) underfill 108-SM GPUs ($1.48$--$1.91$ threads/SM), so randomized batch sweeps ($B \in \{1,\dots,30\}$, zero GPX6 byte change; Figure~\ref{fig:nvcomp}(b), Table~\ref{tab:cross_domain_batch}) test SM scaling across GPU codecs. At $B=2$, host buffer packing dips A100 GPX6 encode from $265.9$ to $212.1$\,MB/s while decode rises to $195.9$\,MB/s. For $B \ge 4$, SM occupancy ($21{,}728$ threads, $201.2$/SM at $B=30$) scales A100 GPX6 to $396.3/418.6$\,MB/s at $B=8$ and $474.1/482.3$\,MB/s ($4.81/5.98$\,GB/s resident) at $B=30$. That exceeds batched A100 \texttt{nvCOMP ANS} ($299.1/288.2$\,MB/s at 64\,KiB, $34.97\%$; $287.7/284.6$ at 8\,KiB), \texttt{nvCOMP Zstd} ($275.0$ at 8\,KiB; $270.2$ at 64\,KiB), and \texttt{MANS -u2} ($130.8/76.5$\,MB/s, $99.15\%$). On mobile 4080, shared CPU/GPU power peaks H2H at $B=16$ ($433.4/485.6$\,MB/s vs.\ $276.7/304.2$ at $B=30$; batched 8\,KiB \texttt{nvCOMP ANS} hits $704.3/738.3$\,MB/s at $40.47\%$), while resident GPX6 scales from $1.47/0.67$ to $3.49/4.97$\,GB/s.

\begin{table}[!htbp]
\centering
\scriptsize
\renewcommand{\arraystretch}{0.75}
\setlength{\tabcolsep}{2.7pt}
\begin{tabular}{llrrrr}\toprule
Domain (Raw Size) & Codec / Configuration & Agg\% & 4080 E/D & A100 E/D & 4080/A100 Res. \\\midrule
Suite~2 CT (47.94\,MB) & GPX6 Full-6 MDL (1D, GPU) & 31.37 & 973.1/657.9 & 326.9/261.8 & 3.64/3.50 \\
Suite~2 CT (47.94\,MB) & OpenZL-9 num16 (1D, MT) & 33.22 & 188.6/789.6 & 76.1/279.3 & --- \\
Suite~2 CT (47.94\,MB) & GPU $\Delta_1$+nvCOMP ANS (1D) & 38.26 & 1037.1/871.1 & 287.0/301.6 & 13.77/45.83 \\
Suite~2 CT (47.94\,MB) & MANS \texttt{-u2} 3D ADM (GPU) & 49.64 & 366.8/182.1 & 190.9/60.3 & --- \\
Suite~2 CT (47.94\,MB) & 2D JPEG-LS (CharLS, MT) & 28.11 & 564.5/479.2 & 206.1/201.7 & --- \\
Suite~2 CT (47.94\,MB) & 2D JPEG~XL (effort-7, MT) & 25.55 & 29.5/214.1 & 17.5/83.7 & --- \\\midrule
Suite~3 EDF (150.60\,MB) & GPX6 Full-6 ($s^*\!=\!1$, GPU) & 47.66 & 464.1/403.6 & 150.7/133.0 & 2.66/2.51 \\
Suite~3 EDF (150.60\,MB) & ZigShuf+Zstd-16 (1D, MT) & 48.27 & 10.6/451.4 & 7.8/197.3 & --- \\
Suite~3 EDF (150.60\,MB) & OpenZL-9 num16 (1D, MT) & 48.92 & 184.5/662.4 & 73.8/249.3 & --- \\\midrule
Suite~4a NIfTI (1.29\,MB) & GPX6 Full-6 MDL (512-row) & 27.04 & 105.3/85.8 & 103.0/58.9 & --- \\
Suite~4a NIfTI (1.29\,MB) & GPX6 256-row (capacity abl.) & 25.96 & 160.0/102.3 & 118.4/66.4 & --- \\
Suite~4a NIfTI (1.29\,MB) & OpenZL-9 num16 (1D, MT) & 23.30 & 42.5/431.6 & 23.8/176.4 & --- \\\midrule
Suite~4b Audio (1.32\,MB) & GPX6 Full-6 MDL (512-row) & 64.49 & 57.6/41.5 & 50.6/28.9 & --- \\
Suite~4b Audio (1.32\,MB) & GPX6 256-row (capacity abl.) & 63.81 & 82.7/49.0 & 58.9/32.3 & --- \\
Suite~4b Audio (1.32\,MB) & FLAC-16 LPC (audio, MT) & 53.65 & 61.5/175.4 & 60.6/87.0 & --- \\\midrule
Suite~1 $B\!=\!1$ (88.90\,MB) & GPX6 batch sweep ($B\!=\!1$) & 29.75 & 300.8/240.6 & 265.9/164.6 & 1.47/1.40 \\
Suite~1 $B\!=\!30$ (88.90\,MB) & GPX6 batched (21,728 thr) & 29.75 & 276.7/304.2 & 474.1/482.3 & 3.49/4.81 \\\bottomrule
\end{tabular}
\caption{Unit-stride Suites~2--4 and paired batch-sweep scaling (H2H \texttt{MB/s}; 4080/A100 Res.\,Enc \texttt{GB/s}).}
\label{tab:cross_domain_batch}
\end{table}

Figures~\ref{fig:nvcomp}--\ref{fig:edf_ct} and Table~\ref{tab:cross_domain_batch} calibrate boundaries. On Sleep-EDF (150.60\,MB; files $\le 51.44$\,MB), Full-6 MDL picks Modes~2--3 on 11/11 tiles (47.66\% vs.\ 48.27\% ZigShuf+Zstd-16, 51.31\% GPU $\Delta_1$+nvCOMP ANS, and 59.19\% GPU $\Delta_1$+\texttt{MANS -u2}; 46.81\% on Combined-10 at 494.1/409.8\,MB/s, $3.67\times$ and $1.91\times$ faster than CPU-MT modeling). On 16-bit CT (47.94\,MB), 1D GPX6 reaches 31.37\% at 973.1/657.9\,MB/s vs.\ 1D OpenZL-9 (33.22\%), 3D \texttt{MANS -u2} (49.64\% GPU, 42.80\% CPU-R), and PNG-16 (38.97\%), while 2D $(H,W)$ codecs reach lower ratios: JPEG-LS (28.11\%), JPEG~2000 (28.09\%), HTJ2K (30.10\%), and JPEG~XL (25.55\%, 29.5\,MB/s). Screened LZ/graph (Screen+\allowbreak\texttt{OpenZL-9} 27.00\%, $24.6$--$25.2$\,MB/s; Screen+\allowbreak\texttt{ZigShuf}+\allowbreak\texttt{Zstd-16} 27.38\%, $1.8$--$3.1$\,MB/s, $p_{\text{Holm}}=1.00$) exploits matches beyond 8\,KiB chunks (11.14\% vs.\ 18.12\% on CEBSDB), and LPC (\texttt{FLAC-16}) leads on smooth $C \le 8$ signals and 1.32\,MB speech (53.65\% vs.\ 64.49\% GPX6). Raw passthrough and lags $s > 32$ remain future work.

\section{Conclusion}
GPX6 pairs microsecond 32-lane CUDA warp lag screening with a threshold-free $3\times 2$ Cartesian byte-bank rANS coder, OpenMP Zstd-1 framing, and a CUDA-free C++ receiver. Across 70 public 16-bit files (446.00\,MB), GPX6 aligns multi-channel lags without headers, lowers ratios vs.\ GPU \texttt{nvCOMP} and 16-bit \texttt{MANS}, scales to 474.1/482.3\,MB/s (4.81/5.98\,GB/s resident) on A100, and maps CPU LZ/graph, LPC, and 2D boundaries.

\Section{References}
\bibliographystyle{ieeetr}
\bibliography{references}

\end{document}